\documentclass[prd,preprint,superscriptaddress]{revtex4}
\usepackage{revsymb}
\usepackage[T1]{fontenc}
\usepackage{indentfirst}
\usepackage{amsmath}
\usepackage{graphicx}
\usepackage{bm}
\usepackage[dvips]{color}
\usepackage{amssymb}
\usepackage{hyperref}%for .pdf
\usepackage{epstopdf}%for .pdf
\usepackage{subfig}
\begin{document}
\title{Baryonic matter perturbations in decaying vacuum cosmology}
\author{R.F. vom Marttens\footnote{E-mail: rodrigovonmarttens@gmail.com}}
\affiliation{Universidade Federal do Esp\'{\i}rito Santo,
Departamento
de F\'{\i}sica\\
Av. Fernando Ferrari, 514, Campus de Goiabeiras, CEP 29075-910,
Vit\'oria, Esp\'{\i}rito Santo, Brazil}

\author{W.S. Hip\'{o}lito-Ricaldi\footnote{E-mail: wiliam.ricaldi@ufes.br}}
\affiliation{Universidade Federal do Esp\'{\i}rito Santo, Departamento de Ci\^encias  Naturais, CEUNES\\
Rodovia BR 101 Norte, km. 60, CEP 29932-540,
S\~ao Mateus, Esp\'{\i}rito Santo, Brazil}
\author{W. Zimdahl\footnote{E-mail: winfried.zimdahl@pq.cnpq.br}}
\affiliation{Universidade Federal do Esp\'{\i}rito Santo,
Departamento
de F\'{\i}sica\\
Av. Fernando Ferrari, 514, Campus de Goiabeiras, CEP 29075-910,
Vit\'oria, Esp\'{\i}rito Santo, Brazil}

%%%%%%%%%%%%%%%%%%%%%%%%%%%%%%%%%%%%%%%%%%%%%%%%%%%%%%%%%%%%%%%%%%%%%%%%%%%%%

\begin{abstract}
We consider the perturbation dynamics for the cosmic baryon fluid and determine the corresponding power spectrum for a $\Lambda(t)$CDM model in which a cosmological term decays into dark matter linearly with the Hubble rate.
The model is tested by a joint analysis of data from supernovae of type Ia (SNIa) (Constitution and Union 2.1),
baryonic acoustic oscillation (BAO), the position of the first peak of the anisotropy spectrum of the cosmic microwave background (CMB) and large-scale-structure (LSS) data (SDSS DR7).
While the homogeneous and isotropic background dynamics is only marginally influenced by the baryons, there are modifications on the perturbative level if a separately conserved baryon fluid is included.
Considering  the present baryon fraction as a free parameter, we reproduce the observed abundance
of the order of $5\%$ independently of the dark-matter abundance which is of the order of
$32\%$ for this model. Generally, the concordance between background and perturbation dynamics is improved if baryons are explicitly taken into account.
\end{abstract}
\date{\today}
\maketitle

%%%%%%%%%%%%%%%%%%%%%%%%%%%%%%%%%%%%%%%%%%%%%%%%%%%%%%%%%%%%%%%%%%%%%%%%%%%%%%

\section{Introduction}
Explaining structure formation in the expanding Universe is one of the major topics in cosmology and astrophysics.
According to the current main-stream understanding, dark matter (DM) and dark energy (DE) are the dynamically dominating components of the Universe \cite{wmap,wmap12,planck}.
Baryons contribute only a small fraction of less than 5\% to the cosmic energy budget.
The standard $\Lambda$CDM model does well in fitting most observational data but there is an ongoing interest in alternative models within and beyond General Relativity.
A class of alternative models within General Relativity "dynamizes" the cosmological constant,  resulting in so-called $\Lambda(t)$CDM models.
Taking the cosmological principle for granted, cosmic structures represent inhomogeneities in the matter distribution on an otherwise spatially homogeneous and isotropic background.
Dynamical DE models, $\Lambda(t)$CDM models are a subclass of them, have to deal with inhomogeneities
of the DE component in addition to the matter inhomogeneities to which they are coupled.
This makes these models technically more complex
than the standard model. Ignoring perturbations of the DE component
altogether may lead to inconsistencies and unreliable conclusions concerning the interpretation of observational data \cite{Park-Hwang}.
Whether or not DE perturbations are relevant has to be decided on a case-by-case basis.
The directly observed inhomogeneities are of baryonic nature.
From the outset it is not clear that the inhomogeneities in the baryonic matter coincide with the inhomogeneities of the DM distribution.
In particular, if DM interacts nongravitationally with DE, which happens in $\Lambda(t)$CDM models, while
baryonic matter is in geodesic motion, this issue has to be clarified.
A reliable description of the observed matter distribution has to consider
the perturbation dynamics of the baryon fraction even though the latter only marginally influences the homogeneous and isotropic cosmic background dynamics.
Then, in models with dynamical DE, the perturbations of baryonic matter will necessarily be coupled to the inhomogeneities of both DM and DE.
In a general context, the importance of including the physics of the baryon component in the cosmic dynamics has been
emphasized recently \cite{nature}.

In this paper we extend a previously established decaying vacuum model \cite{humberto,julio,zimdahl,saulo,saulochap} by including a separately conserved baryon fluid with a four-velocity that differs from the four-velocity of the DM  component.
The basic ingredient of this model is a DE component with an energy density proportional to the Hubble rate.
 Moreover, it is characterized by an equation-of-state (EoS) parameter $-1$ for vacuum. Equivalently, the resulting  dynamics can be understood as a scenario of DM particle production at a constant rate \cite{saulo} or as the dynamics of a non-adiabatic Chaplygin gas \cite{saulochap}.
DE perturbations for this model are explicitly related to DM perturbations and their first derivative with respect to the scale factor in a scale-dependent way. It has been shown that on scales that are relevant for structure formation, DE fluctuations are smaller than the DM  fluctuations by several orders of magnitude \cite{zimdahl}.
Our analysis will be performed within a gauge-invariant formalism in terms of variables adapted to
comoving observers \cite{VDF}.
We shall derive a set of two second-order equations that couple the total fractional energy-density perturbations
of the cosmic medium to the difference between these total perturbations and the fractional baryonic perturbations.
The perturbations of the baryon fluid are then found as a suitable linear combination.

As far as the background dynamics is concerned, our updated tests against observations from SNIa, BAO and the position of the first acoustic peak of the CMB spectrum confirm previous results \cite{Pigozzo2011}.
Including the LSS data improves the concordance of the model compared with the case without a separately conserved baryon component.
The joint analysis allows us to predict the baryon abundance of the Universe independently of the DM abundance. The corresponding probability density function (PDF) exhibits a pronounced peak at about 5\% for this
abundance. This is a new feature which entirely relies on a separate consideration of the baryon fluid.

The paper is organized as follows. In Sec.~\ref{model} we establish the basic relations of our three-component
model of DE, DM and baryons. In Sec.~\ref{background} we recall the homogeneous and isotropic background dynamics of this model. Sec.~\ref{perturbations} is devoted to a gauge-invariant perturbation analysis which results in an explicit expression for the energy-density perturbations of the baryon fluid. In Sec.~\ref{observations} we test the model against observations using both background and LSS data. Our results are summarized in Sec.~\ref{conclusions}.

\section{The model}
\label{model}
We describe the cosmic medium as a perfect fluid with a conserved energy momentum
tensor
\begin{equation}
T_{ik} = \rho u_{i}u_{k} + p h_{ik}\ , \qquad T_{\ ;k}^{ik} = 0\,,
\label{T}
\end{equation}
where $u^{i}$ is the cosmic four-velocity, $h _{ik}=g_{ik} + u_{i}u_{k}$ and $g_{ik}u^{i}u^{k} = -1$. Here, $\rho$ is the energy density for a comoving (with $u^{i}$) observer and $p$ is the fluid pressure.
Latin indices run from $0$ to $3$.
Let us consider a three--component system by
assuming a split of the total energy-momentum tensor in (\ref{T}) into a DM component (subindex M), a DE component (subindex X) and a baryonic component (subindex B),
\begin{equation}\label{Ttot}
T^{ik} = T_{M}^{ik} + T_{X}^{ik} + T_{B}^{ik}\,.
\end{equation}
Each of the components is also modeled as a perfect fluid
with ($A= M, X$, B)
\begin{equation}\label{TA}
T_{A}^{ik} = \rho_{A} u_A^{i} u^{k}_{A} + p_{A} h_{A}^{ik} \
,\qquad\ h_{A}^{ik} = g^{ik} + u_A^{i} u^{k}_{A} \,.
\end{equation}
DM and baryonic matter are assumed to be pressureless.
In general, each component has its own four-velocity with $g_{ik}u_{A}^{i}u_{A}^{k} = -1$.
According to the model to be studied here we include
a (so far unspecified) interaction between the dark components:
\begin{equation}\label{Q}
T_{M\ ;k}^{ik} = Q^{i}\qquad T_{X\ ;k}^{ik} = - Q^{i}\,.
\end{equation}
Then, the energy-balance equations of the dark components are
\begin{equation}
-u_{Mi}T^{ik}_{M\ ;k} = \rho_{M,a}u_{M}^{a} +  \Theta_{M}\rho_{M}  = -u_{Ma}Q^{a}\
\label{eb1}
\end{equation}
and
\begin{equation}
-u_{Xi}T^{ik}_{X\ ;k} = \rho_{X,a}u_{X}^{a} +  \Theta_{X} \left(\rho_{X} + p_{X}\right) = u_{Xa}Q^{a}\,.
\label{eb2}
\end{equation}
The baryonic component is  separately conserved,
\begin{equation}
-u_{Bi}T^{ik}_{B\ ;k} = \rho_{B,a}u_{B}^{a} +  \Theta_{B}\rho_{B} = 0\,.
\label{ebb}
\end{equation}
 The quantities $\Theta_{A}$ are defined as $\Theta_{A} = u^{a}_{A;a}$. For the homogeneous and isotropic background we assume $u_{M}^{a} = u_{X}^{a} = u_{B}^{a} = u^{a}$. Likewise, we have the momentum balances
\begin{equation}
h_{Mi}^{a}T^{ik}_{M\ ;k} = \rho_{M}\dot{u}_{M}^{a}  = h_{M i}^{a} Q^{i}\,,
\label{mb1}
\end{equation}
\begin{equation}
h_{Xi}^{a}T^{ik}_{X\ ;k} = \left(\rho_{X} + p_{X}\right)\dot{u}_{X}^{a} + p_{X,i}h_{X}^{ai} = - h_{X i}^{a} Q^{i}\,,
\label{mb2}
\end{equation}
and
\begin{equation}
h_{Bi}^{a}T^{ik}_{B\ ;k} = \rho_{B}\dot{u}_{B}^{a}  = 0\,,
\label{mbb}
\end{equation}
where $\dot{u}_{A}^{a} \equiv u_{A ;b}^{a}u_{A}^{b}$.
The source term $Q^{i}$ is split into parts proportional and perpendicular to the total four-velocity according to
\begin{equation}
Q^{i} = u^{i}Q + \bar{Q}^{i}\,,
\label{Qdec}
\end{equation}
where $Q = - u_{i}Q^{i}$ and $\bar{Q}^{i} = h^{i}_{a}Q^{a}$ with $u_{i}\bar{Q}^{i} = 0$.
The contribution $T_{X}^{ik}$ is supposed to describe some form of DE. In the simple case of an EoS  $p_{X} = - \rho_{X}$, where $\rho_{X}$ is not necessarily constant, we have
\begin{equation}
T_{X}^{ik} = - \rho_{X}g^{ik}\,.
\label{Tx}
\end{equation}
Dynamically, an energy-momentum tensor like this corresponds to a time-dependent cosmological term.
Various approaches to such type of $\Lambda(t)$  cosmology term can be found in the literature \cite{Lambda(t)}.
Since the only time scale in a homogeneous and isotropic universe is the Hubble time $H^{-1}$, the simplest
phenomenological guess here is $\rho_{X} \propto H$. Interestingly, this guess has some support from particle physics. The QCD vacuum condensate associated to the chiral phase transition leads to a vacuum density proportional to $H$ \cite{QCD}.
It is a dynamics along this line which we intend to study here, albeit in an entirely phenomenological context.
An obvious covariant generalization of a cosmological term that, in the homogeneous and isotropic background, decays linearly with the Hubble rate $H$, i.e., $\rho_{X} \propto H$,
is
\begin{equation}
\rho_{X} = \frac{\sigma}{3}\Theta\,,\qquad p_{X} = - \frac{\sigma}{3}\Theta\,,
\label{rX}
\end{equation}
where $\Theta \equiv u^{a}_{;a}$ is the expansion scalar and  $\sigma$ is a constant. In the homogeneous and isotropic background one has $\Theta = 3H$ and recovers $\rho_{X} \propto H$.

\section{Background dynamics}
\label{background}

The homogeneous and isotropic background dynamics  is governed by Friedmann's equation
\begin{equation}
3 H^{2} = 8\pi G \rho = 8\pi G \left(\rho_{M} + \rho_{X} + \rho_{B}\right)
=  8\pi G \left(\rho_{M} + \rho_{B} + \sigma H\right)\
\label{fried}
\end{equation}
and
\begin{equation}
\dot{H} = - 4\pi G  \left(\rho + p\right) = - 4\pi G \left(\rho_{M} + \rho_{B}\right) \,.
\label{dH}
\end{equation}
Combining Eqs.~(\ref{fried}) and (\ref{dH}) we obtain
\begin{equation}
\dot{H} = - \frac{3}{2}H^{2} +  4\pi G \sigma H\,.
\label{dH1}
\end{equation}
Changing to the scale factor $a$ as independent variable, the solution of Eq.~(\ref{dH1}) is
\begin{equation}
H = \frac{8\pi G }{3}\sigma + \left(H_{0} -  \frac{8\pi G }{3}\sigma\right) a^{-3/2} \,,
\label{Hsol}
\end{equation}
where a subindex 0 indicates the present value of the corresponding quantity and where we put $a_{0} = 1$.
With
\begin{equation}
3H_{0}^{2} = 8\pi G \rho_{0}\,,\quad \Omega_{M0}\equiv \frac{\rho_{M0}}{\rho_{0}}\,,\quad \Omega_{B0}\equiv \frac{\rho_{B0}}{\rho_{0}}\,,\quad\sigma = \frac{\rho_{0}}{H_{0}}\left(1 - \Omega_{M0} - \Omega_{B0}\right)\,,
\label{sig}
\end{equation}
the Hubble rate (\ref{Hsol}) may be written as
\begin{equation}
H = H_{0}\left[1 - \Omega_{M0} - \Omega_{B0} + \left(\Omega_{M0} + \Omega_{B0}\right)a^{-3/2}\right]
\,.
\label{Hom}
\end{equation}
The existence of the last relation in (\ref{sig}) implies that $\sigma$ is not an additional parameter.
The limit of a vanishing $\sigma$ is the Einstein-de Sitter universe, not the $\Lambda$CDM model.
There is no $\Lambda$CDM limit of the dynamics described by the Hubble rate (\ref{Hom}).
The background source terms are
\begin{equation}
u_{a}Q^{a} = - Q = -\dot{p}_{X} =  \sigma\dot{H}\quad \mathrm{and} \quad \bar{Q}^{a} = 0\
\label{Q01}
\end{equation}
and the energy densities $\rho_{M}$ and $\rho_{X}$  are given by
\begin{equation}
\frac{\rho_{M}}{\rho_{0}} = \left(\Omega_{M0} + \Omega_{B0}\right)a^{-3/2}\left[1 - \Omega_{M0} - \Omega_{B0} + \left(\Omega_{M0} + \Omega_{B0} - \frac{\Omega_{B0}}{\Omega_{M0} + \Omega_{B0}} \right)a^{-3/2}\right]
\
\label{rmom}
\end{equation}
and
\begin{equation}
\frac{\rho_{X}}{\rho_{0}} = \left(1 - \Omega_{M0} - \Omega_{B0}\right)\left[1 - \Omega_{M0} - \Omega_{B0} + \left(\Omega_{M0} + \Omega_{B0}\right)a^{-3/2}\right]
\,,
\label{rxom}
\end{equation}
respectively. The baryon energy density is
\begin{equation}
\frac{\rho_{B}}{\rho_{0}} = \Omega_{B0}a^{-3} \,.
\label{rb}
\end{equation}
With~(\ref{Hom}) - ~(\ref{rb}) the background dynamics for the three-component system is exactly solved.
An additional radiation component (subscript R) can be included approximately \cite{saulorad}:
\begin{equation}\label{Happr}
H = H_{0}\left[\left[1 - \Omega_{M0} - \Omega_{B0} + \left(\Omega_{M0} + \Omega_{B0}\right)a^{-3/2}\right]^{2} + \Omega_{R0}a^{-4}\right]^{1/2}
\,.
\end{equation}
(Notice that this is an exact solution of the dynamics only for $\Omega_{R0} = 0$.)
It can be shown that for the standard-model values of $\Omega_{M0}$, $\Omega_{B0}$ and $\Omega_{R0}$ the deviation of (\ref{Happr}) from the exact numerical solution for the Hubble rate is only of the order of 0.6\%.

\section{Perturbations}
\label{perturbations}

\subsection{Balance and conservation equations}

First-order perturbations will be denoted by a hat symbol. While for the background $u_{M}^{a} = u_{B}^{a} = u_{X}^{a} = u^{a}$ is assumed to be valid, the first-order perturbations of these quantities are different, in general.  The perturbed time components of the four-velocities, however, still coincide:
\begin{equation}
\hat{u}_{0} = \hat{u}^{0} = \hat{u}_{M}^{0} =  \hat{u}_{B}^{0} = \hat{u}_{X}^{0}  = \frac{1}{2}\hat{g}_{00}\,.
\label{u0}
\end{equation}
According to the perfect-fluid structure of both the total energy-momentum tensor (\ref{T}) and the energy-momentum tensors of the components in (\ref{TA}), and with $u_{M}^{a} = u_{B}^{a} = u_{X}^{a} = u^{a}$ in the background, we have first-order energy-density perturbations
$\hat{\rho} = \hat{\rho}_{M} + \hat{\rho}_{B} + \hat{\rho}_{X}$, pressure perturbations $\hat{p} = \hat{p}_{M} + \hat{p}_{B} + \hat{p}_{X} = \hat{p}_{X}$
and
\begin{equation}
\hat{T}^{0}_{\alpha} = \hat{T}^{0}_{M\alpha} + \hat{T}^{0}_{B\alpha} +\hat{T}^{0}_{X\alpha}\quad\Rightarrow\quad
\left(\rho + p\right)\hat{u}_{\alpha} = \rho_{M}\hat{u}_{M\alpha} + \rho_{B}\hat{u}_{B\alpha} + \left(\rho_{X} + p_{X}\right)\hat{u}_{X\alpha}
\,.
\label{T0al}
\end{equation}
For $p_{X} = - \rho_{X}$ it follows
\begin{equation}
p_{X} = - \rho_{X} \ \Rightarrow\  \rho + p = \rho_{M} + \rho_{B} \ \Rightarrow\
\hat{u}_{\alpha} = \frac{\rho_{M}}{\rho_{M} + \rho_{B}}\hat{u}_{M\alpha} + \frac{\rho_{B}}{\rho_{M} + \rho_{B}}\hat{u}_{B\alpha}
\,.
\label{ual}
\end{equation}
The perturbations of the time derivatives of the spatial components of the four-velocities differ from the time derivatives of the perturbations by the spatial gradient of $g_{00}$:
\begin{equation}\label{}
\hat{\dot{u}}_{\alpha} = \dot{\hat{u}}_{\alpha} - \frac{1}{2}g_{00,\alpha}\,, \qquad
\hat{\dot{u}}_{M\alpha} = \dot{\hat{u}}_{M\alpha} - \frac{1}{2}g_{00,\alpha}\,, \qquad \hat{\dot{u}}_{B\alpha} = \dot{\hat{u}}_{B\alpha} - \frac{1}{2}g_{00,\alpha}\,.
\end{equation}
The total first-order energy conservation reads
\begin{equation}\label{ebaltot}
\dot{\hat{\rho}}
+ \dot{\hat{\rho}}\hat{u}^{0} + \hat{\Theta}\left(\rho_{M} + \rho_{B}\right)
+ \Theta\left(\hat{\rho} + \hat{p}\right) = 0\,,
\end{equation}
while the separate balances are
\begin{equation}\label{ebalM}
\dot{\hat{\rho}}_{M} + \dot{\rho}_{M}\hat{u}^{0} + \hat{\Theta}_{M}\rho_{M}
+ \Theta\hat{\rho}_{M} = Q = - \left(u_{Ma}Q^{a}\right)^{\hat{}}\,,
\end{equation}
\begin{equation}\label{ebalX}
\dot{\hat{\rho}}_{X} + \dot{\rho}_{X}\hat{u}^{0}
+ \Theta\left(\hat{\rho}_{X} + \hat{p}_{X}\right) = \left(u_{Xa}Q^{a}\right)^{\hat{}}\
\end{equation}
and
\begin{equation}\label{ebalB}
\dot{\hat{\rho}}_{B} + \dot{\rho}_{B}\hat{u}^{0} + \hat{\Theta}_{B}\rho_{B}
+ \Theta\hat{\rho}_{B} = 0\,.
\end{equation}
Comparing the total first-order energy conservation (\ref{ebaltot}) with
the sum of the separate balances (\ref{ebalM}), (\ref{ebalX}) and (\ref{ebalB}) results in
\begin{equation}\label{hTheta}
\hat{\Theta}\left(\rho_{M} + \rho_{B}\right) = \hat{\Theta}_{M}\rho_{M} + \hat{\Theta}_{B}\rho_{B}
+ \left(u_{Ma}Q^{a}\right)^{\hat{}}  - \left(u_{Xa}Q^{a}\right)^{\hat{}}\,.
\end{equation}
To be consistent with the last equation in (\ref{ual}), the last two terms on the right-hand side of (\ref{hTheta})
have to cancel each other. This establishes a relation between the perturbations of the projected interaction terms.

We shall restrict ourselves to scalar perturbations which are described by the line element
\begin{equation}
\mbox{d}s^{2} = - \left(1 + 2 \phi\right)\mbox{d}t^2 + 2 a^2
F_{,\alpha }\mbox{d}t\mbox{d}x^{\alpha} +
a^2\left[\left(1-2\psi\right)\delta _{\alpha \beta} + 2E_{,\alpha
\beta} \right] \mbox{d}x^\alpha\mbox{d}x^\beta \,.\label{ds}
\end{equation}
We also define the three-scalar quantities
 $v$, $v_{M}$ and  $v_{B}$ by
\begin{equation}
a^2\hat{u}^\mu + a^2F_{,\mu} = \hat{u}_\mu \equiv v_{,\mu}\,,\quad a^2\hat{u}_{M}^\mu + a^2F_{,\mu} = \hat{u}_{M\mu} \equiv v_{M,\mu}\,,\quad a^2\hat{u}_{B}^\mu + a^2F_{,\mu} = \hat{u}_{B\mu} \equiv v_{B,\mu}\,.
\label{}
\end{equation}
With the abbreviation
\begin{equation}
\chi \equiv a^2\left(\dot{E} -F\right) \,,
\label{}
\end{equation}
the perturbed scalars $\Theta_{M}$, $\Theta_{B}$ and $\Theta$ are
\begin{equation}
\hat{\Theta}_{M} = \frac{1}{a^2}\left(\Delta v_{M} +\Delta \chi\right) -
3\dot{\psi} - 3 H\phi \ , \quad \hat{\Theta}_{B} = \frac{1}{a^2}\left (\Delta v_{B} +\Delta \chi\right) -
3\dot{\psi} - 3 H\phi \
\label{Thetaexp1}
\end{equation}
and
\begin{equation}
\hat{\Theta} = \frac{1}{a^2}\left (\Delta v +\Delta \chi\right) -
3\dot{\psi} - 3 H\phi\,,\label{Thetaexp}
\end{equation}
respectively, where $\Delta$ denotes the three-dimensional Laplacian.
The last relation of (\ref{ual}) then implies
\begin{equation}\label{}
\left(\rho_{M} + \rho_{B}\right)v = \rho_{M}v_{M} + \rho_{B}v_{B}\,.
\end{equation}
Moreover, as already mentioned, consistency with (\ref{hTheta}) requires
\begin{equation}\label{}
\left(u_{Ma}Q^{a}\right)^{\hat{}}  = \left(u_{Xa}Q^{a}\right)^{\hat{}}\,.
\end{equation}
In terms of the fractional quantities
\begin{equation}\label{}
\delta = \frac{\hat{\rho}}{\rho}\,,\quad \delta_{M} = \frac{\hat{\rho}_{M}}{\rho_{M}}\,,\quad\delta_{X} = \frac{\hat{\rho}_{X}}{\rho_{X}}\,,\quad\delta_{B} = \frac{\hat{\rho}_{B}}{\rho_{B}}\,,
\end{equation}
the energy balances (\ref{ebaltot}), (\ref{ebalM}), (\ref{ebalX}) and (\ref{ebalB}) transform into
\begin{equation}\label{ebaldelta}
\dot{\delta} + \frac{\dot{\rho}}{\rho}\hat{u}^{0} + \hat{\Theta}\frac{\rho_{M} + \rho_{B}}{\rho}
+ \Theta\frac{p}{\rho}\left(\frac{\hat{p}}{p} - \delta\right) = 0\,,
\end{equation}
\begin{equation}\label{ebalMdelta}
\dot{\delta}_{M} + \frac{\dot{\rho}_{M}}{\rho_{M}}\hat{u}^{0} + \hat{\Theta}_{M}
= \frac{\hat{Q}}{\rho_{M}} - \frac{Q}{\rho_{M}}\delta_{M}
\,,
\end{equation}
\begin{equation}\label{ebalXdelta}
\dot{\delta}_{X} + \Theta\left(\frac{\hat{p}_{X}}{\rho_{X}} + \delta_{X}\right)
= \frac{1}{\rho_{X}}\left(u_{Xa}Q^{a}\right)^{\hat{}} + \frac{Q}{\rho_{X}}\left(\delta_{X}
+ \hat{u}^{0}\right)
\
\end{equation}
and
\begin{equation}\label{ebalBdelta}
\dot{\delta}_{B} + \frac{\dot{\rho}_{B}}{\rho_{B}}\hat{u}^{0} + \hat{\Theta}_{B}
= 0
\,,
\end{equation}
respectively.

The total momentum conservation reads (recall that $p_{X} = - \rho_{X}$)
\begin{equation}\label{mbt}
\left(\rho_{M} + \rho_{B}\right)\dot{u}^{a} + p_{X,i}h^{ai} = 0\,.
\end{equation}
The DM and DE momentum balances are given by (\ref{mb1}) and (\ref{mb2}), respectively, with $p_{X} = - \rho_{X}$.
The baryon-fluid motion is geodesic according to (\ref{mbb}).

Our aim is to calculate the energy-density perturbations of the baryon component. In the following subsection we establish, in a first step, an equation for the perturbations of the total energy density. Subsequently, we shall derive an equation for the difference between total and baryonic density perturbations. From the solutions of this system of coupled second-order equations we then obtain the desired
perturbations of the baryon fluid.

\subsection{Perturbations of the total energy density}

To obtain an equation for the total energy-density perturbations
it is convenient to introduce gauge-invariant quantities, adapted to an observer that is comoving with the total fluid four-velocity,
\begin{equation}
\delta^{c} \equiv \delta + \frac{\dot{\rho}}{\rho} v\,, \quad
\hat{\Theta}^{c} \equiv \hat{\Theta} + \dot{\Theta} v\,, \quad  \hat{p}^{c} \equiv \hat{p} + \dot{p}v
\,.
\label{gi}
\end{equation}
Then, the total energy and momentum conservations (\ref{ebaltot}) and (\ref{mbt}), respectively, can be combined into
\begin{equation}\label{balcomb}
\dot{\delta}^{c} - \Theta\frac{p}{\rho}\delta^{c} + \hat{\Theta}^{c}\left(1+\frac{p}{\rho}\right) = 0\,.
\end{equation}
The perturbation $\hat{\Theta}$ has to be determined from the Raychaudhuri equation
\begin{equation}
\dot{\Theta} + \frac{1}{3}\Theta^{2} - \dot{u}^{a}_{;a} + 4\pi G \left(\rho + 3
p\right) = 0\,,\label{Ray}
\end{equation}
where we have neglected shear and vorticity.
At first order we have
\begin{equation}\label{dThetacfin}
\dot{\hat{\Theta}}^{c} + \frac{2}{3}\Theta\hat{\Theta}^{c} + 4\pi G\rho\delta^{c}
+ \frac{1}{a^2}\frac{\Delta \hat{p}^{c}}{\rho + p} = 0\,.
\end{equation}
Combining Eqs.~(\ref{balcomb}) and (\ref{dThetacfin}) and changing to $a$ as independent variable ($\delta^{c\prime} \equiv \frac{d \delta^{c}}{d a}$),
we obtain
\begin{equation}
\delta^{c\prime\prime} + \left[\frac{3}{2}-\frac{15}{2}\frac{p}{\rho}+ 3\frac{\dot{p}}{\dot{\rho}}\right]\frac{\delta^{c\prime}}{a}
- \left[\frac{3}{2} + 12\frac{p}{\rho} - \frac{9}{2}\frac{p^{2}}{\rho^{2}} - 9\frac{\dot{p}}{\dot{\rho}}
\right]\frac{\delta^{c}}{a^{2}}
+ \frac{k^{2}}{a^{2}H^{2}}\frac{\hat{p}^{c}}{\rho a^{2}}
= 0\,,
  \label{dddeltak}
\end{equation}
where $k$ is the comoving wavenumber.
According to (\ref{rX}), for the present model
\begin{equation}\label{}
\hat{p}^{c} = - \frac{\sigma}{3}\hat{\Theta}^{c}
\end{equation}
is valid.
With the help of
(\ref{balcomb}) we find that the pressure perturbation is not just proportional to the energy-density perturbation but to the derivative of $\delta^{c}$ as well:
\begin{equation}\label{hatpc1}
\hat{p}^{c} = - \frac{1}{3}\frac{p}{1 + \frac{p}{\rho}}
\left[a\delta^{c\prime} - 3\frac{p}{\rho}\delta^{c}\right]\,.
\end{equation}
For the later important gauge-invariant combination $\hat{p}^{c} - \frac{\dot{p}}{\dot{\rho}}\rho\delta^{c}$
we have
\begin{equation}\label{nad}
\hat{p}_{nad} \equiv \hat{p} - \frac{\dot{p}}{\dot{\rho}}\rho\delta = \hat{p}^{c} - \frac{\dot{p}}{\dot{\rho}}\rho\delta^{c} =
- \frac{1}{3}\frac{p}{1 + w}
\left[a\delta^{c\prime} + \frac{3}{2}\left(1 - w\right)\delta^{c}\right]\,.
\end{equation}
This quantity describes the non-adiabatic pressure perturbations.

With the expression (\ref{hatpc1}) for the pressure perturbations, Eq.~(\ref{dddeltak}) takes the final form
\begin{equation}
% \nonumber to remove numbering (before each equation)
\delta^{c\prime\prime} + \left[\frac{3}{2}- 6w - \frac{1}{3}
\frac{w}{1+ w}\frac{k^{2}}{a^{2}H^{2}}\right]\frac{\delta^{c\prime}}{a}
- \left[\frac{3}{2} + \frac{15}{2}w - \frac{9}{2}w^{2}
- \frac{w^{2}}{1+ w}\frac{k^{2}}{a^{2}H^{2}}\right]\frac{\delta^{c}}{a^{2}}
= 0\,.
  \label{dddeltak1}
\end{equation}
Here, the total EoS parameter $w = \frac{p}{\rho}$ is explicitly given by
\begin{equation}\label{w(a)}
w = \frac{p}{\rho} =  - \frac{\sigma H}{\rho} = - \frac{1}{1 + r a^{-3/2}}\,,
\end{equation}
where
\begin{equation}\label{}
r \equiv \frac{\Omega_{M0} + \Omega_{B0}}{1 - \Omega_{M0} - \Omega_{B0}}\
\end{equation}
is the present-time ratio of total matter (DM and baryonic matter) to DE.
It is remarkable that there appears a scale-dependence in the $\delta^{c\prime}$ term in Eq.~(\ref{dddeltak1}).
A similar feature holds in bulk-viscous models which are characterized by a non-adiabatic dynamics as well \cite{VDF}.

At high redshifts with $a\ll 1$ the EoS parameter $w$ tends to zero and (\ref{dddeltak1}) approaches
\begin{equation}
\delta^{c\prime\prime} + \frac{3}{2}\frac{\delta^{c\prime}}{a}
- \frac{3}{2}\frac{\delta^{c}}{a^{2}}
= 0 \qquad (a \ll 1)\,,
  \label{prpreds}
\end{equation}
i.e., we recover the equation for density perturbations in an Einstein-de Sitter universe.

\subsection{Relative energy-density perturbations}

As already mentioned, we shall calculate
the baryonic matter perturbations via the total energy-density perturbations, governed by Eq.~(\ref{dddeltak1}), and the relative energy perturbations
$\frac{\hat{\rho}}{\rho + p} - \frac{\hat{\rho}_{B}}{\rho_{B}}$. It is the dynamics of this difference which we shall consider
in the present subsection.
Let us consider to this purpose equations (\ref{ebaldelta}) and (\ref{ebalBdelta}). In (\ref{ebaldelta}) we introduce
\begin{equation}\label{}
D\equiv \frac{\hat{\rho}}{\rho + p}\quad\Rightarrow \quad \delta = D\left(1+\frac{p}{\rho}\right)\,,
\end{equation}
in terms of which Eq.~(\ref{ebaldelta}) reads
\begin{equation}\label{dotD}
\dot{D} + \Theta \left(\frac{\hat{p}}{\rho + p} - \frac{\dot{p}}{\dot{\rho}}D\right)
+ \hat{\Theta} - \Theta\hat{u}^{0} = 0\,.
\end{equation}
Combining the conservation equation (\ref{dotD}) for the total energy with
the energy conservation (\ref{ebalBdelta}) of the baryons and defining $S_{B} \equiv D - \delta_{B}$, we obtain
\begin{equation}\label{D-}
\dot{S}_{B} + \left(\hat{\Theta} - \hat{\Theta}_{B}\right)
+ \Theta \left(\frac{\hat{p}}{\rho + p} - \frac{\dot{p}}{\dot{\rho}}D\right)
= 0\,.
\end{equation}
In the following we shall derive an equation for $S_{B}$ in which this quantity is coupled to the total energy-density perturbations $\delta^{c}$. While the physical meaning of  $\delta^{c}$ is obvious, the situation
seems less clear for $S_{B}$. Simply from the definition one has
\begin{equation}\label{}
S_{B}
= \frac{\rho_{X}}{\rho_{M} + \rho_{B}}\delta_{X} + \frac{\rho_{M}}{\rho_{M} + \rho_{B}}\left(\delta_{M} - \delta_{B}\right)\,.
\end{equation}
If the DE perturbations can be neglected, which is the case in many situations (cf. \cite{zimdahl}), one has $S_{B} \propto \delta_{M} - \delta_{B}$.
Thus it represents a measure for the difference in the fractional perturbations of DM and
baryonic matter. It is useful
as an auxiliary quantity since both the total energy-momentum and the baryon energy-momentum are conserved.

According to the expressions (\ref{Thetaexp1}) and (\ref{Thetaexp}) the difference between the quantities $\hat{\Theta}$ and  $\hat{\Theta}_{B}$ is
\begin{equation}\label{hatT-}
\hat{\Theta} - \hat{\Theta}_{B} = \frac{1}{a^{2}}\Delta\left(v - v_{B}\right)\,.
\end{equation}
Differentiating equation (\ref{D-}) and using the definition of $\hat{p}_{nad}$ in (\ref{nad}) results in
\begin{equation}\label{D-.}
\ddot{S}_{B} + \left(\hat{\Theta} - \hat{\Theta}_{B}\right)^{\displaystyle \cdot}
+ \left[\Theta \frac{\hat{p}_{nad}}{\rho + p}\right]^{\displaystyle \cdot} = 0\,.
\end{equation}
To deal with the time-derivative of expression (\ref{hatT-}) we consider the momentum conservations (\ref{mbt}) and (\ref{mbb})
 which, at first order, can be written as
\begin{equation}\label{}
\dot{v} + \phi = - \frac{\hat{p}^{c}}{\rho + p}\ \quad \mathrm{and } \quad \dot{v}_{B} + \phi = 0\,,
\end{equation}
respectively.
It follows that
\begin{equation}\label{diffv}
\left(v - v_{B}\right)^{\displaystyle \cdot} = - \frac{\hat{p}^{c}}{\rho + p}\,.
\end{equation}
With (\ref{diffv}) and (\ref{hatT-})
the resulting $k$-space equation for $S_{B}$ is
\begin{equation}\label{D-..k}
 \ddot{S}_{B}+ 2 H \dot{S}_{B}  +
\frac{k^{2}}{a^{2}}\frac{\hat{p}^{c}}{\rho + p}
+ \left[3H \frac{\hat{p}^{c}_{nad}}{\rho + p}\right]^{\displaystyle \cdot}
+ 6H^{2}\frac{\hat{p}^{c}_{nad}}{\rho + p} = 0\,.
\end{equation}
Introducing the explicit expressions
(\ref{hatpc1}) and (\ref{nad}), use of (\ref{dddeltak1}) to eliminate the second derivative of $\delta^{c}$ provides us with
\begin{eqnarray}
% \nonumber to remove numbering (before each equation)
S_{B}^{\prime\prime} + \frac{3}{2}\left(1-w\right)\frac{S_{B}^{\prime}}{a}&=& \frac{w}{\left(1 + w\right)^{2}} \left[
\left(3 + \frac{3}{2}w + \frac{1}{3}\frac{1 + 2w}{1+w}\frac{k^{2}}{a^{2}H^{2}}\right)\frac{\delta^{c\prime}}{a}\qquad\qquad\qquad
\right. \nonumber \\
&&\left. \qquad\qquad + \left(\frac{9}{2} - \frac{9}{4}w - \frac{9}{4}w^{2} - w\frac{1 + 2w}{1+w}\frac{k^{2}}{a^{2}H^{2}}\right)\frac{\delta^{c}}{a^{2}}\right]\,.
\label{Sprpr}
\end{eqnarray}
The total density perturbation $\delta^{c}$ and its first derivative appear as inhomogeneities in the equation for $S_{B}$.
Eqs.~(\ref{dddeltak1}) and (\ref{Sprpr}) are the key equations of this paper.
In the next section we demonstrate how a solution of the coupled system (\ref{dddeltak1}) and (\ref{Sprpr}) will allow us to obtain the perturbations of the baryon fluid.

It is expedient to notice that for $a \ll 1$ one has $w\approx 0$ and the total cosmic medium behaves as dust. Under this condition
the right-hand side of Eq.~(\ref{Sprpr}) vanishes and we can use
$S_{B} = $ const $\approx 0$ as initial condition for the numerical analysis.

\subsection{Baryonic energy-density perturbations}

By definition, the fractional baryonic energy-density perturbations $\delta_{B} = \frac{\hat{\rho}_{B}}{\rho_{B}}$ are determined by $D$ and $S_{B}$,
\begin{equation}\label{}
\delta_{B} = D - S_{B}\,.
\end{equation}
Since $S_{B}$ is gauge-invariant by itself, we may write
\begin{equation}\label{}
S_{B} = D - \frac{\hat{\rho}_{B}}{\rho_{B}} = \frac{\delta^{c}}{1+w} - \delta_{B}^{c}\,,
\end{equation}
where
\begin{equation}\label{}
\delta_{B}^{c} = \delta_{B} + \frac{\dot{\rho}_{B}}{\rho_{B}}v = \delta_{B} - \Theta v\,.
\end{equation}
Consequently, the comoving (with $v$) baryon energy-density perturbations are given by the combination
\begin{equation}\label{deltaB}
\delta_{B}^{c} = \frac{\delta^{c}}{1+w} - S_{B}\,.
\end{equation}

%%%%%%%%%%%%%%%%%%%%%%%%%%%%%%%%%%%%%%%%%%%%%%%%%%%%%%%%%%%%%%%%%%%%%%%%%%%%%%%%%%%%%%%%%%%%%%%%%%%%%%%%%%%%%%%%%%

\begin{figure}[!htb]
\subfloat[]{
\includegraphics[width=7cm]{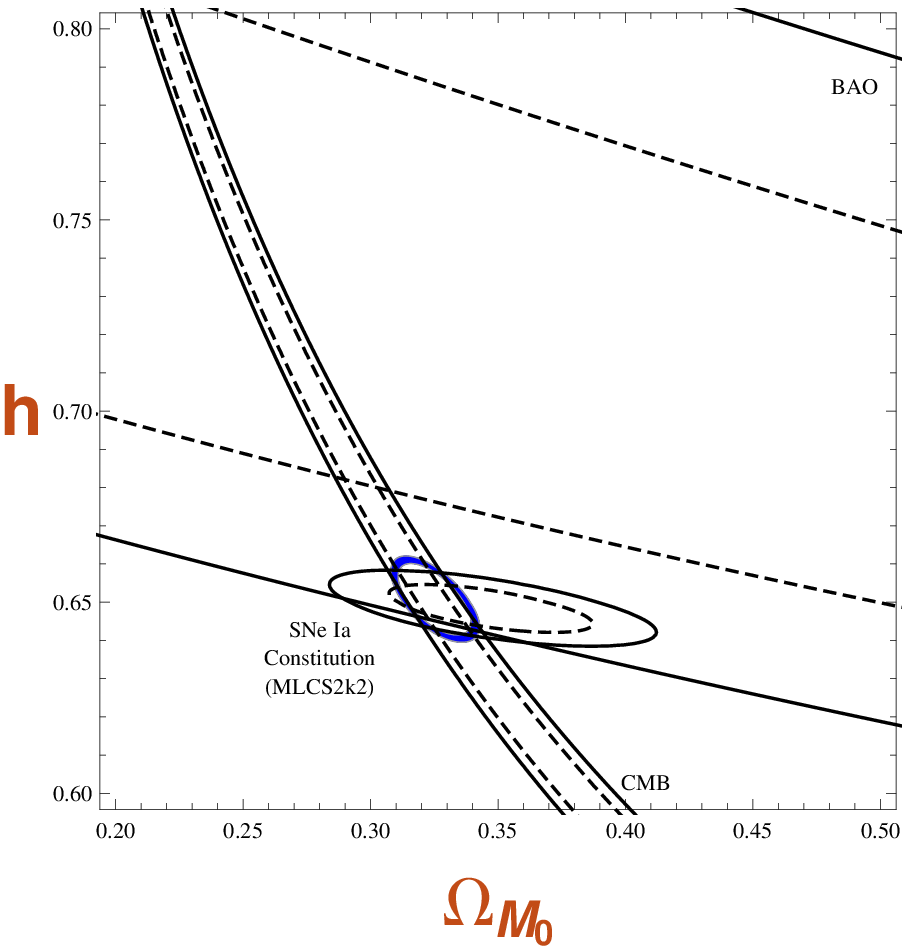}
\label{fig1a}
}
\quad %espaco separador
\subfloat[]{
\includegraphics[width=7cm]{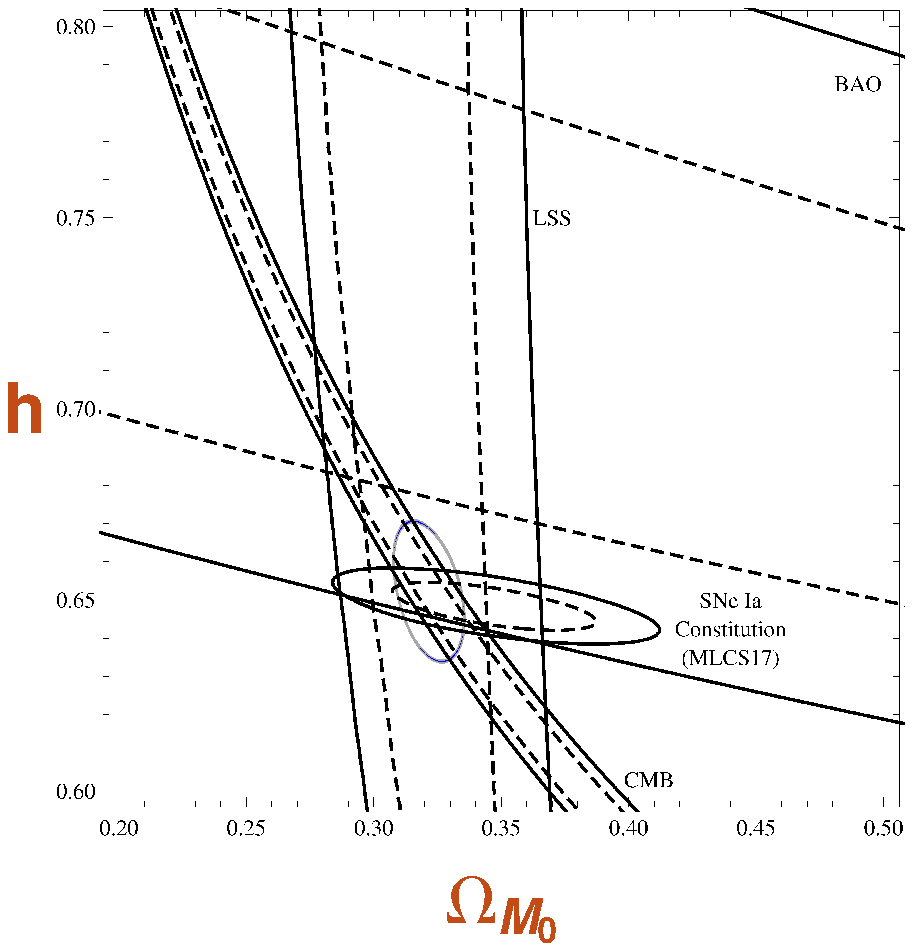}
\label{fig1b}
}
\caption{(a) Constitution data set with  MLCS17 fitter combined with BAO and the position of the first acoustic peak.
(b) The same as in (a) with LSS data added. The dashed and continuous contour lines  refer to the $1\sigma$ and $2\sigma$ confidence regions, respectively. The blue regions indicate the results of the joint tests at the $2\sigma$ level.}
\label{fig1}
\end{figure}

It seems more convenient, however, to consider the perturbations of the baryon fluid with respect
to the velocity potential $v_{B}$ of the baryon component itself.
These perturbations are
obtained via
\begin{equation}\label{vcB}
\delta_{B}^{c_{B}} \equiv \delta_{B} - \Theta v_{B} = \delta_{B}^{c} + \Theta\left(v - v_{B}\right)\,.
\end{equation}
Use of (\ref{D-}) with (\ref{nad}) and (\ref{hatT-}) leads to
\begin{equation}\label{}
\frac{k^{2}}{a^{2}}\left(v - v_{B}\right) = \dot{S}_{B} + 3H \frac{\hat{p}_{nad}}{\rho + p}\,.
\end{equation}
For $\delta_{B}^{c_{B}}$ we obtain
\begin{equation}\label{ccB}
\delta_{B}^{c_{B}} = \delta_{B}^{c} + 3 \frac{a^{2}H^{2}}{k^{2}}\left[aS_{B}^{\prime}
+ \frac{\hat{p}_{nad}}{\rho + p}\right]\,.
\end{equation}
Equation (\ref{ccB}) establishes a relation between perturbations measured by an observer, comoving with
the baryon fluid and perturbations measured by an observer, comoving with the total velocity of the cosmic substratum. Obviously, the difference between both quantities depends on the perturbation scale.
On small scales $\frac{a^{2}H^{2}}{k^{2}} \ll 1$ one has
$\delta_{B}^{c_{B}} \approx \delta_{B}^{c}$, i.e., the difference is negligible.
Explicitly, $\delta_{B}^{c_{B}}$ is given in terms of $\delta^{c}$  and $S_{B}$ and their first derivatives by
\begin{equation}\label{deltacB}
\delta_{B}^{c_{B}} = \frac{\delta^{c}}{1+w} - S_{B}
+ 3 \frac{a^{2}H^{2}}{k^{2}}\left[aS_{B}^{\prime}
- \frac{w}{3}\frac{1}{\left(1+w\right)^{2}}
\left(a\delta^{c\prime} + \frac{3}{2}\left(1-w\right)\delta^{c}\right)\right]\,.
\end{equation}
One has to solve now Eq.~(\ref{dddeltak1}) for $\delta^{c}$ and afterwards equation (\ref{Sprpr}) for
$S_{B}$, in which $\delta^{c}$ and its first derivative appear as inhomogeneities. The coefficients are given by (\ref{Hom}) and (\ref{w(a)}). The initial conditions at high redshift are determined by the Einstein - de Sitter type behavior (\ref{prpreds}) with $S_{B} \approx 0$, equivalent to an almost adiabatic behavior.
The perturbations of the baryonic component then are found by the combinations (\ref{deltaB}) or (\ref{deltacB}).
As already mentioned, because of the factor $\frac{a^{2}H^{2}}{k^{2}}$ in front of the last term on the right-hand side of (\ref{deltacB}) one expects negligible differences between $\delta_{B}^{c_{B}}$ and $\delta_{B}^{c}$ on sub-horizon scales $k^{2} \ll a^{2}H^{2}$.
\begin{figure}[!h]
\subfloat[]{
\includegraphics[width=7cm]{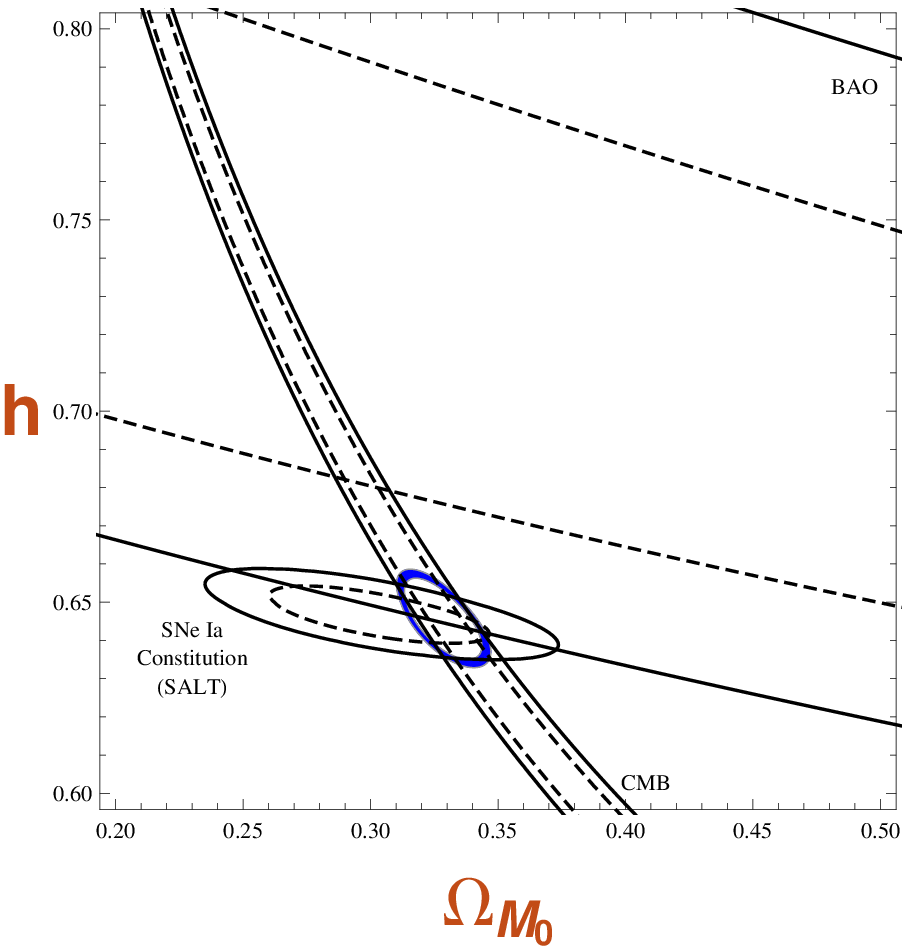}
\label{fig2a}
}
\quad %espaco separador
\subfloat[]{
\includegraphics[width=7cm]{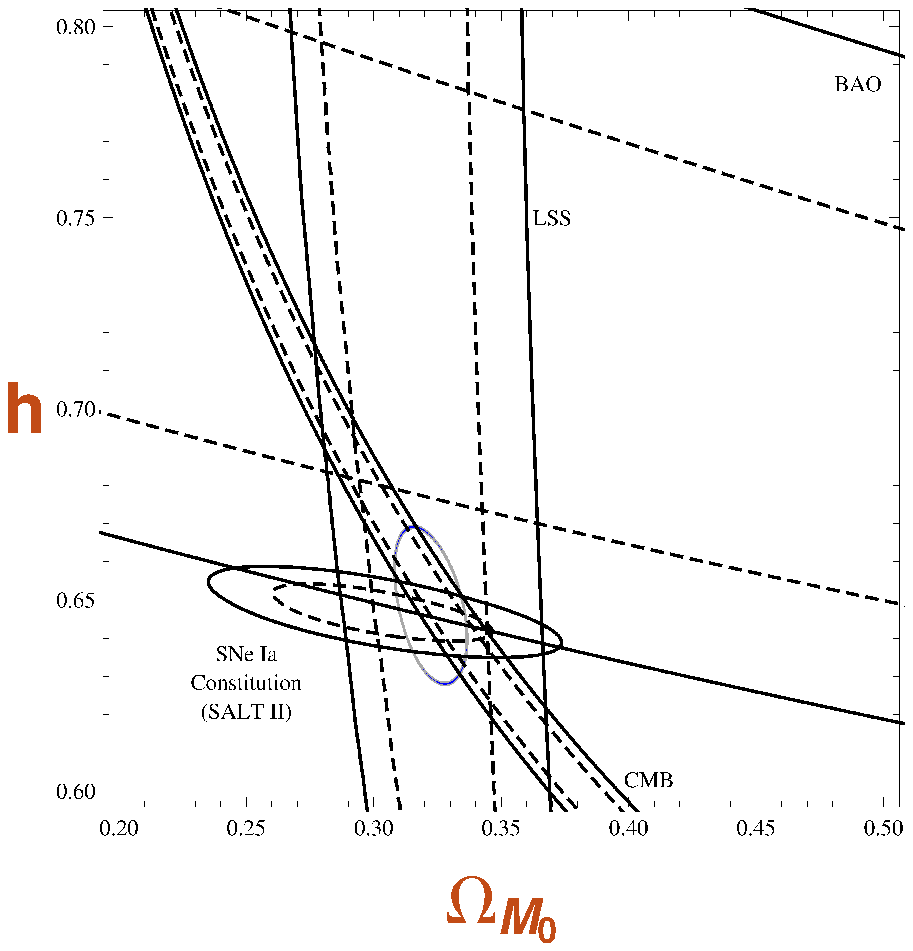}
\label{fig2b}
}
\caption{Data as in Fig.~1, here with SALT II fitter.}
\label{fig2}
\end{figure}

%%%%%%%%%%%%%%%%%%%%%%%%%%%%%%%%%%%%%%%%%%%%%%%%%%%%%%%%%%%%%%%%%%%%%%%%%%%%%%%%%%%%%%%%%%%%%%%%%%%%%%%%%%%%%%%%%
\section{Observational analysis}
\label{observations}
%%%%%%%%%%%%%%%%%%%%%%%%%%%%%%%%%%%%%%%%%%%%%%%%%%%%%%%%%%%%%%%%%%%%%%%%%%%%%%%%%%%%%%%%%%%%%%%%%%%%%%%%%%%%%%%%%%
%%%%%%%%%%%%%%%%%%%%%%%%%%%%%%%%%%%%%%%%%%%%%%%%%%%%%%%%%%%%%%%%%%%%%%%%%%%%%%%%%%%%%%%%%%%%%%%%%%%%%%%%%%%%%%%
As far as the background dynamics is concerned, the explicit inclusion of a baryon component does not significantly change the Hubble rate (\ref{Hom}). It is only the combination $\Omega_{M0} + \Omega_{B0}$ which matters.
For our background tests, which in part are updates of previous studies, we have considered data from SNIa (Constitution \cite{Hicken}
and Union 2.1 \cite{suzuki}), BAO \cite{eisenstein,tegmark,percival} and the position of the first acoustic peak of the CMB spectrum
\cite{hinshaw,spergel}. For a more complete analysis of the SNIa samples and to test the robustness
of the results, we use both the fitters Multicolor Light Curve Shapes (MLCS) \cite{riess1} and Spectral
Adaptive Lightcurve Template (SALT II) \cite{guy,guy1}.

As is well known, SNIa tests are using the luminosity distance modulus
\begin{equation}
\mu=5\log d_L(z)+\mu_0 \
\label{modulo}
\end{equation}
with $\mu_0=42.384-5\log h $, where
\begin{eqnarray}
d_{L}=\left(z+1\right)H_0\int_{0}^{z}\frac{dz'}{H\left(z'\right)}\,
\label{luminosidade}
\end{eqnarray}
and $h$ is given by $H_0 = 100  h \mathrm{km s^{-1} Mpc^{-1}}$.
Tests against BAO data are based on the geometric quantity \cite{eisenstein,tegmark,percival}
\begin{equation}
 D_{v}\left(z\right)=\left[\left(1+z\right)^{2}d_{A}^{2}\frac{z}{H\left(z\right)}\right]^{1/3} z\,,
 \end{equation}
where $d_{A}$ is the angular-diameter distance.
Concerning the position of the first acoustic peak of the CMB anisotropy spectrum, we rely on the distance scale
\cite{hu1,sethi},
%%%%%%%%%%%%%%%%%%%%%%%%%%%%%%%%%%%%%%%%%%%%%%%%%%%%%%%%%%%%%%%%%%%%%%%%%%%%%%%%%%%%%%%%%%%%%%%%%%%%%%%%%%%%%%%5
%%%%%%%%%%%%%%%%%%%%%%%%%%%%%%%%%%%%%%%%%%%%%%%%%%%%%%%%%%%%%%%%%%%%%%%%%%%%%%%%%%%%%%%%%%%%%%%%%%%%%%%%%%%%%%%%%

\begin{figure}[!t]
\subfloat[]{
\includegraphics[width=7cm]{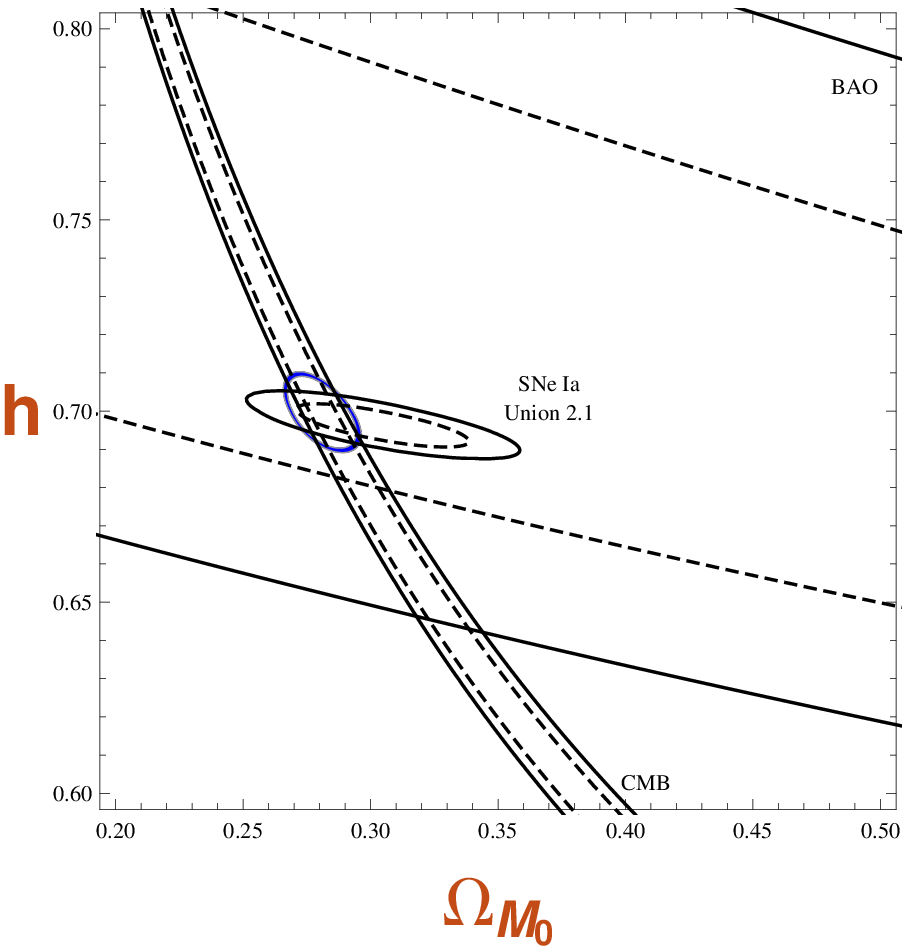}
\label{fig3a}
}
\quad %espaco separador
\subfloat[]{
\includegraphics[width=7cm]{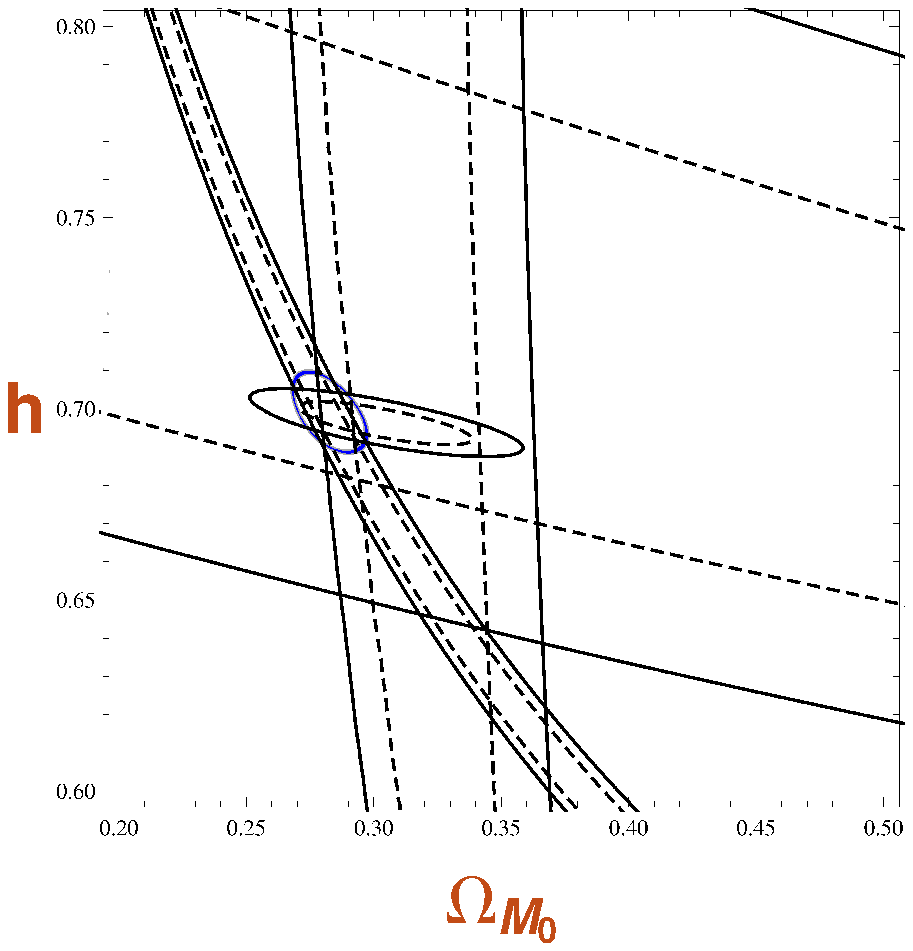}
\label{fig3b}
}
\caption{(a) Union 2.1 data set with SALT II fitter combined with BAO and the position of the first acoustic peak.
(b) The same as in (a) with LSS data added. The dashed and continuous contour lines  refer to the $1\sigma$ and $2\sigma$ confidence regions, respectively. The blue regions indicate the results of the joint tests at the $2\sigma$ level.}
\label{fig3}
\end{figure}
%%%%%%%%%%%%%%%%%%%%%%%%%%%%%%%%%%%%%%%%%%%%%%%%%%%%%%%%%%%%%%%%%%%%%%%%%%%%%%%%%%%%%%%%%%%%%%%%%%%%%%%%%%%%%%%%%%%%

\begin{equation}
l_{1}=l_{A}\left(1-\delta_{1}\right)\,.
\end{equation}
Here, $l_{A}$ is the acoustic scale ($c_{s}$ is the sound speed)
\begin{eqnarray}
l_{A}=\pi \frac{\int\frac{dz}{H(z)}}{\int c_{s}^{2}\frac{dz}{H(z)}}, \qquad  c_{s}^{2}=\sqrt{3+\frac{9\Omega_{B0}}{4\Omega_{R0}}z^{-1}}\,
\end{eqnarray}
and $\delta_{1}\approx 0.267\left(\frac{10\Omega_{R_{0}}}{3\Omega^{2}_{m_{0}}}\right)^{1/10}$ is a correction term,
adapted to the decaying vacuum model \cite{Pigozzo2011}. At the perturbative level we consider the LSS data of Ref.~\cite{sdss} and calculate the baryonic power spectrum $P_k\propto |\delta_B|^2$.

%%%%%%%%%%%%%%%%%%%%%%%%%%%%%%%%%%%%%%%%%%%%%%%%%%%%%%%%%%%%%%%%%%%%%%%%%%%%%%%%%%%%%%%%%%%%%%%%%%%%%%%%%%%%%%%%%%%%%
%%%%%%%%%%%%%%%%%%%%%%%%%%%%%%%%%%%%%%%%%%%%%%%%%%%%%%%%%%%%%%%%%%%%%%%%%%%%%%%%%%%%%%%%%%%%%%%%%%%%%%%%%%%%%%%%%%%
\begin{figure}[!h]
\subfloat[]{
\includegraphics[width=6.5cm]{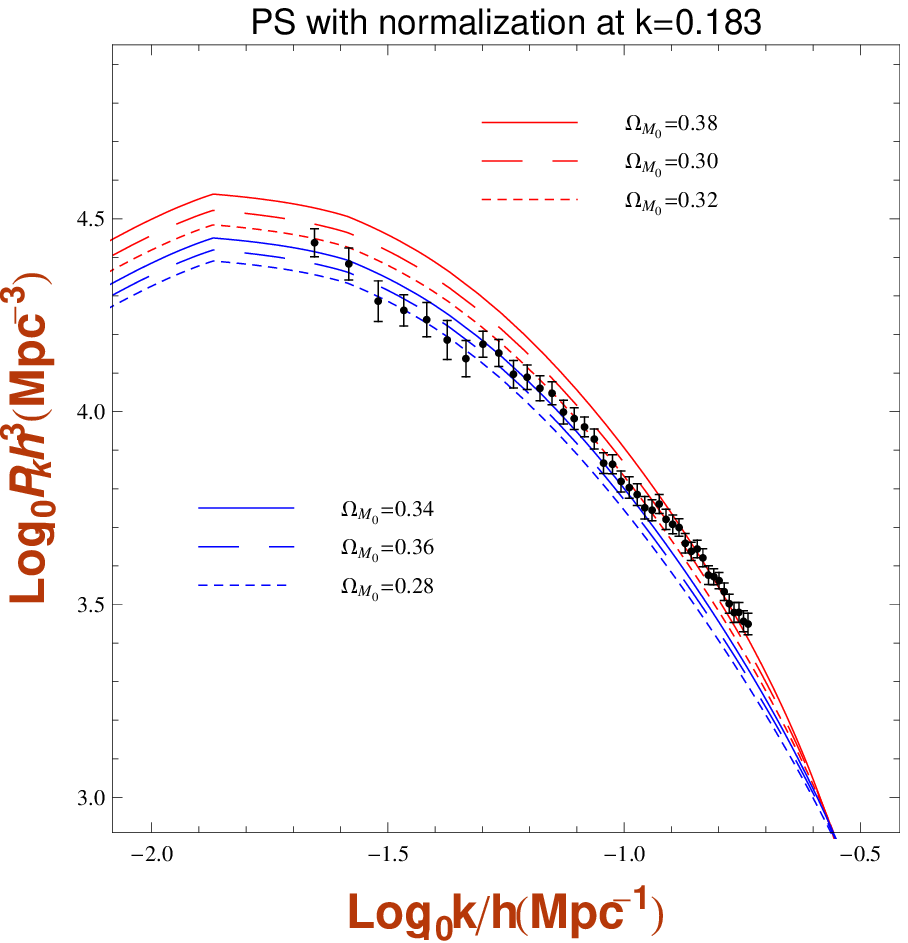}
\label{fig4a}
}
\subfloat[]{
\includegraphics[width=6.5cm]{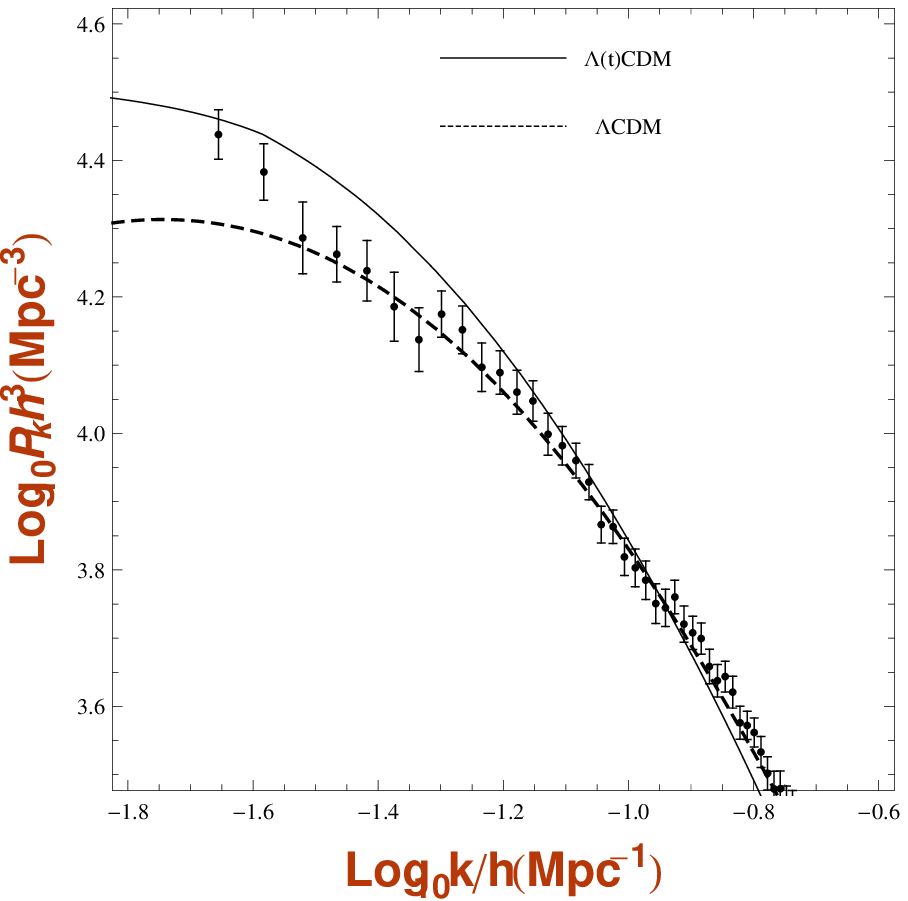}
\label{fig4b}
}
\caption{Left panel: baryonic matter-power spectrum with different values of $\Omega_{M0}$.
Values between $0.28$ and $0.36$ are in reasonable agreement with the LSS data (SDSS DR7).
Notice that these values are considerably lower than those found in \cite{Pigozzo2011,zimdahl,saulochap} ($\sim 0.37 -0.43$)
without a separate baryon component. 
Right panel: best-fit power spectra for the $\Lambda(t)$CDM and $\Lambda$CDM models.}
\label{fig4}
\end{figure}

%%%%%%%%%%%%%%%%%%%%%%%%%%%%%%%%%%%%%%%%%%%%%%%%%%%%%%%%%%%%%%%%%%%%%%%%%%%%%%%%%%%%%%%%%%%%%%%%%%%%%%%%%%%%%%%%%%%%%
%%%%%%%%%%%%%%%%%%%%%%%%%%%%%%%%%%%%%%%%%%%%%%%%%%%%%%%%%%%%%%%%%%%%%%%%%%%%%%%%%%%%%%%%%%%%%%%%%%%%%%%%%%%%%%%%%%%%
\noindent
For our tests we perform a $\chi^2$ analysis, using
\begin{eqnarray}
\chi^{2}(\theta)=\sum\limits_{i=1}^{N}\frac{\left[y_{i}-y\left(x_{i}\vert\theta\right)\right]^{2}}{\sigma_{i}^{2}}\,.
\label{chi2}
\end{eqnarray}
Here, the $y_{i}$ are the observational data (SNIa, CMB, BAO, LSS)  which are compared with the theoretical predictions $y(x_{i}\vert\theta)$, where $\theta$ represents a set of model parameters and $\sigma_i$ denotes the error bars.
Out of $\chi^{2}$  in (\ref{chi2}) one defines the probability distribution function (PDF)
$\mathcal{P}\propto\exp\left(-\frac{\chi^{2}(\theta)}{2}\right)$.  For the present model the set of parameters is
$\theta=(h, \Omega_{B0}, \Omega_{M0})$.
In a first step, however, we fix the baryon
abundance in agreement with primordial nucleosynthesis.
Under this condition the free parameters are the same as in the $\Lambda$CDM model, namely $h$ and the
DM abundance $\Omega_{M0}$.
Our results are presented in figures 1 - 3. The dashed and continuous contour lines in all these figures refer to the
$1\sigma$ and $2\sigma$ confidence levels (CL), respectively.
Fig. 1(a) shows the $h$ - $\Omega_{M_{0}}$ plane based on the Constitution data with MLCS17 fitter
combined  with data from BAO and the position of the first acoustic peak of the CMB. In Fig. 1(b)
we have added LSS data to the background tests of Fig.~1(a). In both cases blue regions mark the results of the joint tests at the $2\sigma$ CL.
Figures 2(a) and 2(b) visualize the
$h$ - $\Omega_{M_{0}}$ plane for the same data as in Figs. 1(a) and 1(b), but with SALT II fitter.
In Figs. 3(a)  and 3(b) the corresponding curves for the
Union 2.1 sample are presented.  Again, in both  cases blue regions
indicate the results of the joint tests at 2$\sigma$ CL. Our background tests largely
reproduce previous results \cite{Pigozzo2011}.  Only that our value for the position of the first acoustic peak
differs slightly from the result of \cite{Pigozzo2011}.
In our case the baryon abundance is fixed both in the Hubble rate and in the expression for the sound speed, in
\cite{Pigozzo2011} it is fixed only for calculating the sound speed.
The best-fit values for the background tests alone are summarized in Table I where we compare our model with the
$\Lambda$CDM model via their $\chi_{\nu}^{2}$ values (reduced $\chi^{2}$ values). For the joint background and LSS tests we find the best-fit values in Table II.

\begin{table}
\caption{Best fit values at the 2$\sigma$ CL using background tests (SNIa, BAO, CMB).}
{\scriptsize{\begin{tabular}{c|ccc|ccc} \hline \hline
&\multicolumn{3}{c|}{\textbf{$\Lambda$CDM} } & \multicolumn{3}{|c}{\textbf{$\Lambda(t)$CDM}} \\\hline \hline
 Test &h&$\Omega_{m0}$&$\chi^2_{\nu}$ & h&$\Omega_{m0}$&$\chi^2_{\nu}$  \\ \hline
 SNIa Constitution (MLCS17) &$0.650^{+0.009}_{-0.009}$&$\,\, 0.324^{+0.056}_{-0.054}$&$\,\,1.087$ & $0.648^{+0.009}_{-0.010}$&$\,\,0.399^{+0.066}_{-0.062}$&$\,\,1.087$  \\ \hline
 SNIa Constitution (SALT II) & $0.649^{+0.010}_{-0.009}$&$0.282^{+0.057}_{-0.060}$ &$0.979$ & $0.647^{+0.011}_{-0.012}$& $0.355^{+0.072}_{-0.066}$& $0.983$ \\ \hline
 SNIa Union 2.1 & $0.700^{+0.008}_{-0.008}$&$0.278^{+0.032}_{-0.040}$ & $0.973$& $0.697^{+0.009}_{-0.009}$& $0.348^{+0.041}_{-0.051}$&$0.975$ \\ \hline
 BAO+CMB+SNIa Constitution (MLCS17)&  $0.656^{+0.010}_{-0.011}$ & $0.255^{+0.015}_{-0.012}$& $1.090$&$0.651^{+0.011}_{-0.016}$ &$0.377^{+0.017}_{-0.018}$&$1.094$\\ \hline
BAO+CMB+SNIa Constitution (SALT II)&  $0.652^{+0.012}_{-0.014}$ & $0.257^{+0.013}_{-0.012}$ &$0.992$ &$0.645^{+0.016}_{-0.019}$ &$0.382^{+0.018}_{-0.018}$ &$0.996$\\ \hline
BAO+CMB+SNeIa Union 2.1&   $0.701^{+0.008}_{-0.007}$ &$0.242^{+0.009}_{-0.008}$  &$0.969$ &  $0.699^{+0.007}_{-0.014}$&$0.328^{+0.013}_{-0.016}$  &$0.974$\\ \hline
\end{tabular}}}
\label{table1}
\end{table}

%%%%%%%%%%%%%%%%%%%%%%%%%%%%%%%%%%%%%%%%%%%%%%%%%%%%%%%%%%%%%%%%%%%%%%%%%%%%%%%%%%%%%%%%%%%%%%%%%%%%%%%%%%%%%%%%%%
\begin{figure}[!b]
\begin{center}
\includegraphics[width=1.0\textwidth]{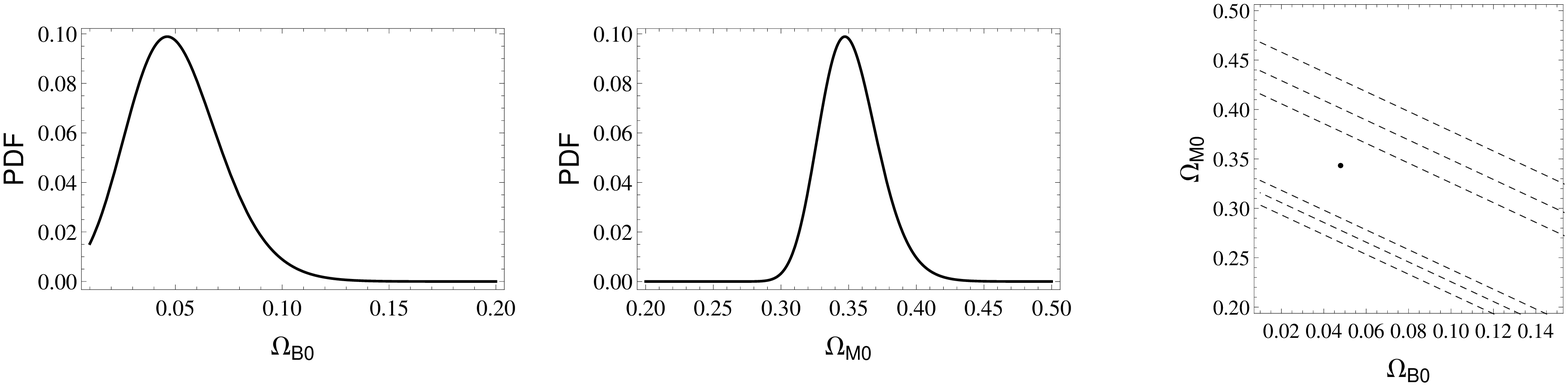}
%\label{fig1}	
\caption{PDFs for the baryon fraction $\Omega_{B0}$ (left panel) and the DM fraction $\Omega_{M0}$ (central panel) based on the LSS data.
The right panel shows the $\Omega_{B0}$-$\Omega_{M0}$ plane with the $1\sigma$, $2\sigma$ and $3\sigma$  contour lines.
The dot indicates the best-fit values $\Omega_{B0}= 0.05\pm 0.02$ and $\Omega_{M0}=0.35\pm0.03$ at the
$2\sigma$ CL.}
\end{center}
\label{fig5}
\end{figure}

Our analysis confirms that the decaying $\Lambda$ model predicts a higher value of the current DM abundance than the
$\Lambda$CDM model. Interestingly, from the LSS data alone we find (at the $2\sigma$ CL) $\Omega_{M0}=0.32\pm 0.04$,
a lower value than for
the model without a separate baryon component \cite{Pigozzo2011,zimdahl,saulochap}, although still higher than in the
$\Lambda$CDM model. The $\chi^2_{\nu}$  values  in Table I  reveal that, as far as the background dynamics is concerned, our
$\Lambda(t)$CDM model is competitive with the $\Lambda$CDM model.
On the other hand, comparing the results for the baryon power spectrum, the situation changes.
While for the data from the 2dFGRS project \cite{cole} we find  $\chi_{\nu}^{2} \approx 0.91$ for the $\Lambda(t)$CDM model
and $\chi_{\nu}^{2} \approx 0.96$ for $\Lambda$CDM, the SDSS DR7 data with their much smaller error bars clearly favor
the $\Lambda$CDM model with $\chi_{\nu}^{2} \approx 0.93$ compared with  $\chi_{\nu}^{2} = 3.63$ of the decaying
$\Lambda$ model. 
The left panel of Fig.~4 visualizes the baryonic power spectrum confronted
with the SDSS DR7 data for different values of $\Omega_{M0}$. The best-fit power spectra for both models are shown in Fig. 4.
One should keep in mind here that in obtaining the spectrum the BBKS transfer function \cite{bbks} was used which naturally favors the $\Lambda$CDM model.

%for different values of $\Omega_{M0}$.

\begin{figure}[!t]
\subfloat[]{
\includegraphics[width=5.5cm]{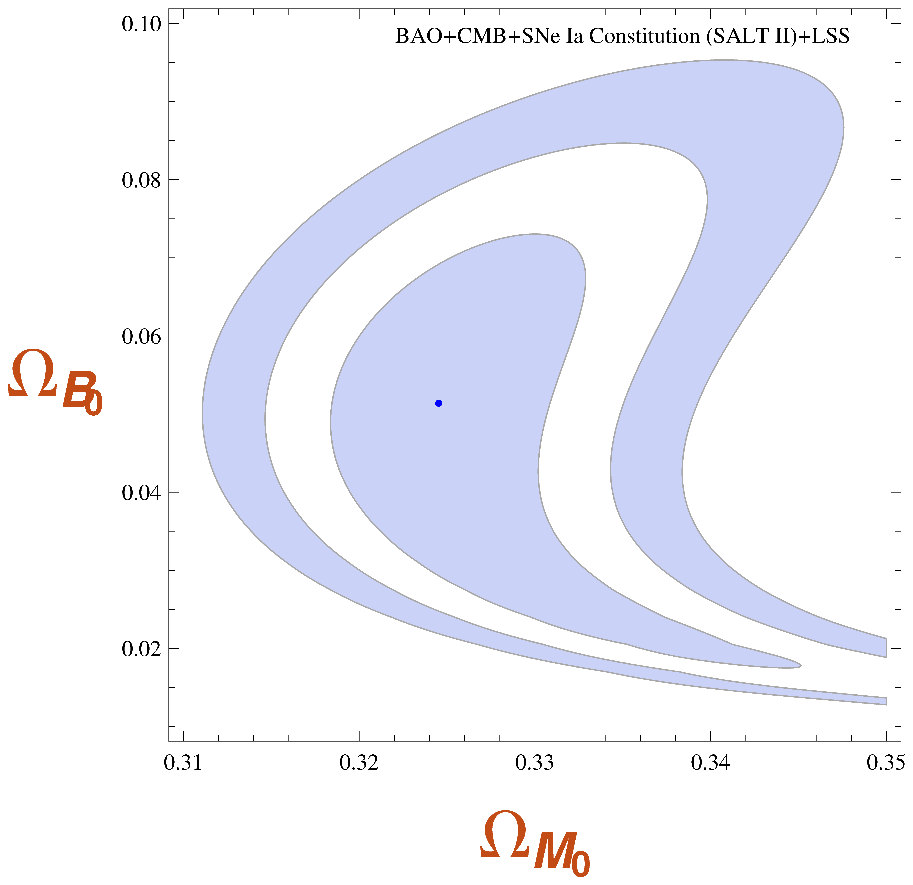}
\label{fig6a}
}
%\quad %espaco separador
\subfloat[]{
\includegraphics[width=5.5cm]{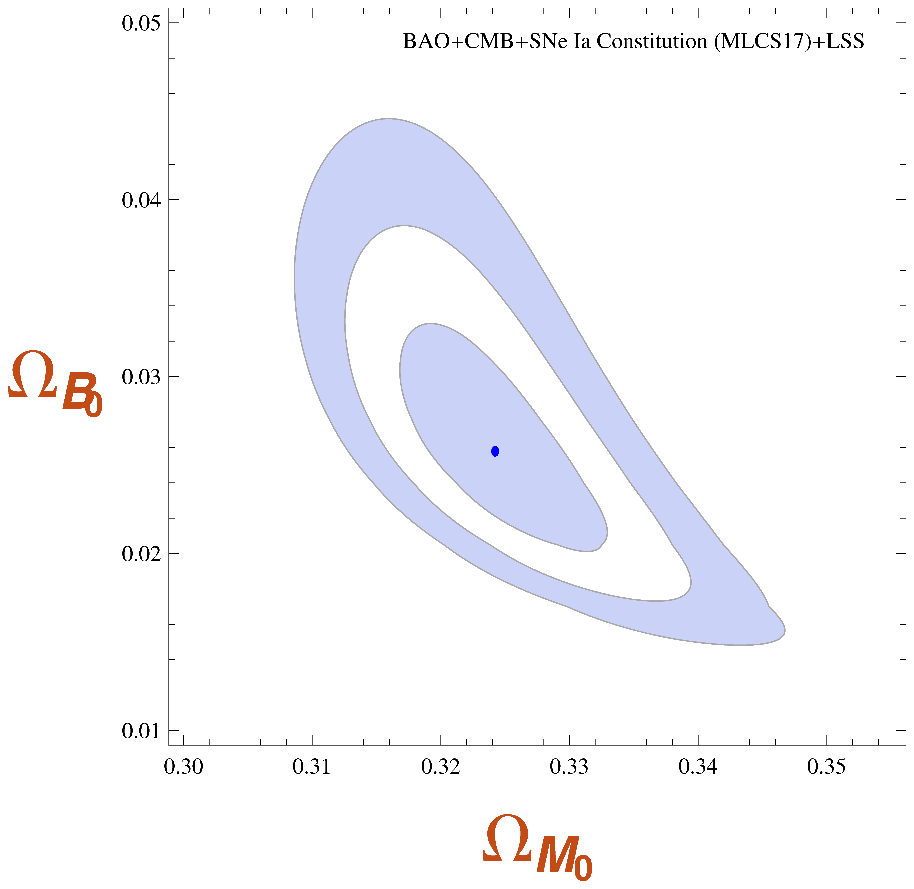}
\label{fig6b}
}
\subfloat[]{
\includegraphics[width=5.5cm]{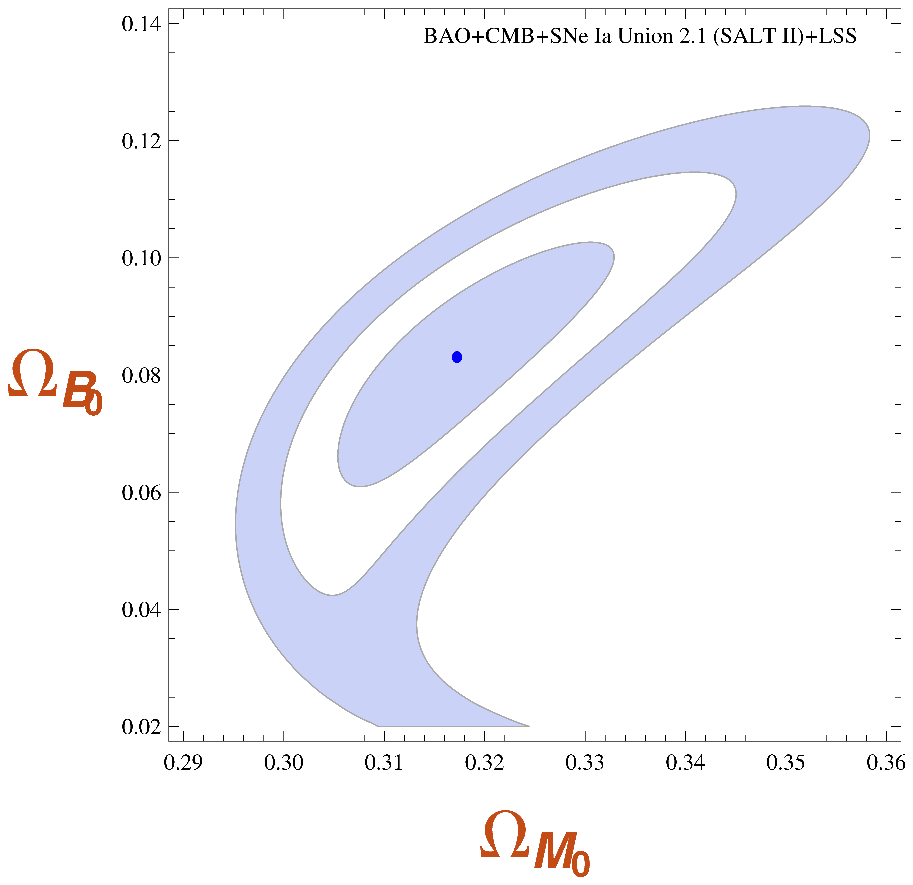}
\label{fig6c}
}
\caption{Two-dimensional contour plots for the abundances of baryons and DM.
(a) Joint analysis with data from LSS, CMB, BAO and Constitution SNIa data with SALT II fitter.
(b) Same data as in (a)  with  MLSCk2 fitter.
(c) Joint analysis with  data from LSS, CMB, BAO  and Union 2.1 SNIa  data. The results for the baryon abundance
(see Table III) are in agreement with primordial nucleosynthesis.
}
\label{fig6}
\end{figure}

In the tests so far the baryon fraction $\Omega_{B0}$ was assumed to be given. Now we relax this assumption
and consider $\Omega_{B0}$ and $\Omega_{M0}$ to be free parameters.
Performing a statistical analysis of the LSS data with $h=0.7$ as a prior (in concordance with our result for the Union2.1 based background test in Tab. I), we obtain the
the two-dimensional curves in the right panel of Fig.~5 with the best-fit values $\Omega_{B0}= 0.05\pm 0.02$ ($2\sigma$ CL) and
$\Omega_{M0}= 0.35\pm 0.03$ ($2\sigma$ CL). The one-dimensional PDF for $\Omega_{B0}$ (left panel of Fig.~5) is then found by fixing
$\Omega_{M0}= 0.35$, the corresponding plot for $\Omega_{M0}$ (central panel) by fixing $\Omega_{B0}= 0.05$.
The same PDFs follow
for a prior  $h=0.65$, indicating that these results do not depend strongly on the specific choice of the prior.
Remarkably, the best-fit value $\Omega_{B0}= 0.05\pm 0.02$ ($2\sigma$ CL) is found to be in agreement with the
result from nucleosynthesis and, at the same time, also demonstrates the consistency of
our approach.
In a next step we performed an enlarged analysis using the entire set of data (SNIa, CMB, BAO and LSS).
This enlarged analysis (see Fig.~6) confirms the LSS-based results of Fig.~5.
The left panel of Fig.~6 shows the two-dimensional contour plots resulting  from a joint test with LSS, CMB, BAO and the Constitution data with SALT II fitter.
Figure~6(b) was obtained with the same data but now with MLSC17 fitter.
On the basis of the Union 2.1 data we found the results in Fig.~6(c).
The best-fit values for the baryon and DM abundances are summarized in Table III.
We conclude that our results for the baryon abundance are in agreement with the results from nucleosynthesis at the 2$\sigma$ CL.
The consistent reproduction of the cosmic baryon abundance
on the basis of data from LSS and background tests is a main achievement of this paper.
\ \\
%%%%%%%%%%%%%%%%%%%%%%%%%%%%%%%%%%%%%%%%%%%%%%%%%%%%%%%%%%%%%%%%%%%%%%%%%%%%%%%%%%%%%%%%%%%%%%%%%%%%%%%%%
{\scriptsize{
\begin{table}
\caption{Best fit values at the 2$\sigma$ CL using joint tests (SNIa, BAO, CMB, LSS).}
{\scriptsize{\begin{tabular}{c|cc|cc} \hline \hline
&\multicolumn{2}{c|}{\textbf{$\Lambda$CDM} } & \multicolumn{2}{|c}{\textbf{$\Lambda(t)$CDM}} \\\hline \hline
 Test &$\Omega_{M0}$&$\chi^2_{\nu}$ & $\Omega_{M0}$&$\chi^2_{\nu}$  \\ \hline
LSS& $0.292^{+0.025}_{-0.023}$& $0.929$&$0.363^{+0.032}_{-0.031}$&$3.634$\\ \hline
BAO+CMB+SNIa Constitution (MLCS17)+LSS &$\,\, 0.315^{+0.026}_{-0.024}$&$\,\,0.970$ &$\,\,0.375^{+0.034}_{-0.036}$&$\,\,1.352$  \\ \hline
BAO+CMB+SNIa Constitution (SALT II)+LSS&$0.310^{+0.025}_{-0.022}$ &$0.975$ & $0.375^{+0.038}_{-0.040}$& $1.352$ \\ \hline
BAO+CMB+SNIa Union 2.1+LSS &$0.284^{+0.021}_{-0.021}$ & $0.963$&  $0.330^{+0.032}_{-0.035}$&$1.240$ \\ \hline
\end{tabular}}}
\label{table2}
\end{table} 
}}
%%%%%%%%%%%%%%%%%%%%%%%%%%%%%%%%%%%%%%%%%%%%%%%%%%%%%%%%%%%%%%%%%%%%%%%%%%%%%%%%%%%%%%%%%%%%%%%%%%%%%%%%%%%%%%%%%%

%%%%%%%%%%%%%%%%%%%%%%%%%%%%%%%%%%%%%%%%%%%%%%%%%%%%%%%%%%%%%%%%%%%%%%%%%%%%%%%%%%%%%%%%%%%%%%%%%%%%%%%%%%%%%%%%%%%%
{\scriptsize{
\begin{table}
\caption{Best-fit values for the $\Lambda$(t)CDM model at the 2$\sigma$ CL using  data from SNIa, CMB, BAO and LSS, considering DM and baryon abundances as free parameters.}
\begin{tabular}{c|cc} \hline \hline
 Test &$\Omega_{B_{0}}$ &$\Omega_{M_{0}}$  \\ \hline\hline
LSS& $0.054^{+0.023}_{-0.018}$  & $0.347^{+0.023}_{-0.025}$  \\ \hline
BAO+CMB+SNe Ia Constitution (MLCS17)+LSS& $0.026^{+0.013}_{-0.008}$ &  $0.324^{+0.015}_{-0.012}$ \\ \hline
BAO+CMB+SNe Ia Constitution (SALT II)+LSS& $0.051^{+0.010}_{-0.010}$ &  $0.325^{+0.015}_{-0.010}$ \\ \hline
BAO+CMB+SNe Ia Union 2.1 (SALT II)+LSS& $0.083^{+0.032}_{-0.041}$ & $0.317^{+0.028}_{-0.018}$ \\ \hline
\end{tabular}
\label{tabela3}
\end{table}
}}

%%%%%%%%%%%%%%%%%%%%%%%%%%%%%%%%%%%%%%%%%%%%%%%%%%%%%%%%%%%%%%%%%%%%%%%%%%%%%%%%%%%%%%%%%%%%%%%%%%%%%%%%%%%%%%
%%%%%%%%%%%%%%%%%%%%%%%%%%%%%%%%%%%%%%%%%%%%%%%%%%%%%%%%%%%%%%%%%%%%%%%%%%%%%%%%%%%%%%%%%%%%%%%%%%%%%%%%%%%%%%%%%%%%

\section{Conclusions}
\label{conclusions}

The components of the cosmological dark sector, DM and DE, are dominating the overall dynamics of the Universe. The small baryonic fraction of presently less than 5\% of the energy budget
does only marginally influence the homogeneous and isotropic expansion history.
With the help of data from SNIa, BAO and the position of the first peak of the CMB anisotropy spectrum
we updated and confirmed previous results for the background.
But as far as structure formation is concerned, the situation is  different.
The directly observed inhomogeneous matter distribution in the Universe is the distribution of visible, i.e., baryonic matter.
While the standard scenario according to which the baryons after radiation decoupling are falling into the potential wells created by the DM inhomogeneities may suggest a similar distribution of DM and baryonic matter, the situation less clear if DM is in (non-gravitational) interaction with DE, while the (directly) observed baryon component is separately conserved.
We have carried out a detailed gauge-invariant perturbation analysis for the baryon fluid in a $\Lambda(t)$CDM cosmology in which a cosmological term is decaying into DM linearly with the Hubble rate.
Our key result is an expression for the fractional baryon energy-density perturbation for an observer comoving with the baryon fluid. Using the LSS data of the SDSS DR7 project we obtained the PDF for the baryon abundance of the Universe independently of the DM abundance. The best-fit value of this abundance is  $\Omega_{B0} = 0.05 \pm 0.02$ (2$\sigma$ CL) in remarkable
agreement with the result from primordial nucleosnthesis. A combined analysis, including also data from SNIa,
BAO and CMB confirms this result.
For the best-fit value of the DM abundance we found
$\Omega_{M0}=0.32\pm0.02$ ($2\sigma$ CL) from the combined analysis (LSS+BAO+SNIa(Union2.1)+CMB) and  $\Omega_{M0}= 0.35\pm 0.03$ ($2\sigma$ CL) from the LSS data alone.
These values are higher than those for the standard model but smaller than the corresponding value for a $\Lambda(t)$CDM model without a separately conserved baryon component. Generally, the explicit inclusion of the baryon fluid improves the concordance between background and perturbation dynamics.
Our results indicate that the investigated $\Lambda(t)$CDM cosmology, which does not have
a $\Lambda$CDM limit, has a competitive background dynamics but as far as the baryon matter power spectrum is concerned, the $\Lambda$CDM model is clearly favored.

%%%%%%%%%%%%%%%%%%%%%%%%%%%%%%%%%%%%%%%%%%%%%%%%%%%%%%%%%%%%%%%%%%%%%%%%%%%%%%%%%%%%%%%%%%%%%%%%%%%%%%%%%%%%%%
%%%%%%%%%%%%%%%%%%%%%%%%%%%%%%%%%%%%%%%%%%%%%%%%%%%%%%%%%%%%%%%%%%%%%%%%%%%%%%%%%%%%%%%%%%%%%%%%%%%%%%%%%%%%%%%%%%%%

\acknowledgments{We thank Saulo Carneiro and J\'{u}lio Fabris for helpful discussions.
Financial support by CAPES, FAPES and CNPq is gratefully acknowledged.  WSHR is thankful to FAPES for the grant (BPC No 476/2013),
under which this work was carried out.}

\end{document}